\documentclass{elsarticle}

\usepackage{amsmath}
\usepackage{amssymb}
\usepackage{graphicx}
\usepackage{booktabs}
\usepackage{hyperref}
\usepackage[numbers]{natbib}
\usepackage{tikz}
\usetikzlibrary{positioning,arrows.meta}
\usepackage{multirow}

\journal{Biomedical Signal Processing and Control}

\begin{document}


\begin{frontmatter}

\title{Decodable but Not Accessible: Auditing Distance-Based Reliability Estimation on Disentangled Skin-Lesion Representations}

\author[inst1]{Duc-Vinh Tran\corref{cor1}}
\ead{vinh.td225538@sis.hust.edu.vn}

\address[inst1]{School of Information and Communication Technology, Hanoi University of Science and Technology, 1 Dai Co Viet Road, Hai Ba Trung, Hanoi 100000, Vietnam}

\cortext[cor1]{Corresponding author.}

\begin{abstract}
%

Distance-based reliability estimation assumes that a representation's
geometry reflects its trustworthiness, yet this assumption is rarely tested
under training interventions that reshape representation geometry directly.
We audit this assumption under domain-adversarial representation learning, using a disentanglement dose-response ladder. Three
checkpoint families share the same architecture and a 16-dimensional
representation, differing only in orthogonality strength
($\lambda_{orth} = 0, 1, 5$). Representation geometry changed
substantially with disentanglement strength: condition number shifted by
two orders of magnitude (Kendall's $\tau=0.84$, exact $p=2.8\times10^{-5}$).
This geometric change was not accompanied by any improvement in reliability
estimation: Mahalanobis-distance AUROC (ISIC-test vs.\ PAD-UFES) remained
flat and below chance ($\approx 0.40$) at every disentanglement level, with
no significant association to any of five geometry metrics tested. The same
failure was shared by cosine-to-centroid and pooled $k$-nearest-neighbor
scorers, as well as three additional, non-distance-based scorers (an
energy-based confidence score, Virtual-Logit Matching, and a kernel density
estimator) -- eight scorers spanning distinct scoring principles, seven of
which converged on the same result; one (the energy-based score) showed an
isolated upward trend with disentanglement strength that we report but do
not treat as evidence against the overall pattern. A supervised probe with no access to the
training objective nonetheless recovered domain membership from the
identical embeddings at 0.72--0.81 AUROC across every disentanglement level,
showing the relevant information was not absent from the representation.
Collectively, these findings indicate that evaluating a learned representation by
classification performance alone can overlook whether the information it
contains is organized in a form downstream reliability estimators can
actually use. Information can remain decodable while becoming largely inaccessible
to non-probing reliability estimators.

\end{abstract}


\end{frontmatter}

\section{Introduction}
\label{sec:introduction}

Deep neural networks are typically evaluated by predictive accuracy, yet deployment safety also depends on the structure of the representation a network learns internally, not only on how it classifies a fixed test set. Whether downstream reliability estimation succeeds therefore depends not only on what a representation encodes, but also on how that information is organized. A representation's geometry---how tightly points cluster by class, how well-conditioned the resulting covariance is, how closely that structure resembles the class-conditional Gaussian shape many downstream estimators assume---is itself shaped by training, and can change substantially without moving classification accuracy at all.

Distance-based reliability and out-of-distribution (OOD) estimation methods are built directly on this geometry. Mahalanobis distance \citep{lee2018simple} scores a test point by its distance to the nearest class-conditional mean under an estimated covariance; cosine-similarity-to-centroid variants replace the covariance-weighted distance with a simpler angular one; non-parametric $k$-nearest-neighbor scorers \citep{sun2022out} drop the notion of a class centroid altogether and use local density in the representation space instead. Despite spanning fully parametric to fully non-parametric, all three rest on the same premise: whatever geometric structure a representation has, a sufficiently faithful measure of distance within it will track how typical, and therefore how trustworthy, a given input is. Other reliability estimators relax this purely geometric premise---Energy \citep{liu2020energy} and Virtual-Logit Matching \citep{wang2022vim} incorporate the classifier's own logits directly, and density-based estimators replace a single Gaussian assumption with a non-parametric one---but share the same underlying constraint: all are computed from a fixed, already-trained representation, without retraining or access to the original objective.

This premise ties reliability estimation to representation geometry by construction, so any training procedure that reshapes representation geometry also reshapes, whether or not this is intended, the reliability estimator built on top of it. Domain-adversarial and orthogonality-based disentanglement training \citep{ganin2015unsupervised} are procedures designed explicitly to reorganize this geometry---separating or suppressing structure associated with a nuisance variable such as acquisition site or imaging domain---and are increasingly used in clinical imaging pipelines for exactly that reason. Whether the geometric changes such training induces are the kind distance-based reliability estimation can still read, or the kind that severs the link between geometry and reliability without affecting classification accuracy, has not, to our knowledge, been audited directly.

We audit this assumption using a disentanglement dose-response ladder that isolates the effect of orthogonality strength while holding the architecture and representation fixed. This design allows us to ask whether geometric change, distance-based reliability estimation, and information decodability remain coupled under disentanglement. Doing so lets us distinguish two claims the existing OOD literature does not, to our knowledge, separate for a training-induced geometric shift: that a representation has lost the information a reliability estimator would need, versus that the representation still holds this information in a form distance-based geometry cannot use.

Across the ladder, representation geometry changes substantially, whereas eight structurally distinct reliability estimators---five distance-based, three not---consistently remain near or below chance ($\sim$0.40 AUROC), with a single exception discussed as a hypothesis-generating observation rather than a counter-finding (Section~\ref{sec:discussion}). A supervised probe applied to the identical representations, in contrast, recovers the relevant domain signal at 0.73--0.81 AUROC. These results are obtained within a domain-adversarial training regime evaluated against its own adversarial target domain (Section~\ref{sec:datasets}), and are scoped accordingly throughout this paper. Information can remain decodable while becoming largely inaccessible to non-probing reliability estimators.

\section{Methods}
\label{sec:methods}

\subsection{Overall study design}
\label{sec:study-design}

\begin{figure}[htbp]
\centering
\begin{tikzpicture}[
    box/.style={draw, rounded corners, align=center, minimum height=1.1cm, minimum width=3.1cm, font=\small},
    arrow/.style={-{Latex[length=2mm]}, thick},
    node distance=0.7cm and 0.5cm
]
\node[box] (ladder) {Disentanglement ladder\\$\lambda_{orth}=0,1,5$ (13 checkpoints)};
\node[box, below=of ladder] (embed) {$z_{lesion}$ embeddings (16-d)\\ISIC-train / ISIC-test / PAD-UFES};

\node[box, below left=1.1cm and -0.9cm of embed] (geom) {Geometry metrics\\(condition number,\\Fisher ratio, Mardia $\kappa$)};
\node[box, below=1.1cm of embed] (scorers) {Distance-based scorers\\(Mahalanobis, cosine,\\pooled $k$-NN)};
\node[box, below right=1.1cm and -0.9cm of embed] (probe) {Domain probe\\(logistic regression,\\linear SVM, random forest)};

\node[box, below=1.3cm of scorers, minimum width=6.5cm] (assoc) {Association across the ladder\\(exact-permutation Kendall's $\tau$ / Jonckheere--Terpstra)};

\draw[arrow] (ladder) -- (embed);
\draw[arrow] (embed) -- (geom);
\draw[arrow] (embed) -- (scorers);
\draw[arrow] (embed) -- (probe);
\draw[arrow] (geom) -- (assoc);
\draw[arrow] (scorers) -- (assoc);
\draw[arrow] (probe) -- (assoc);
\end{tikzpicture}
\caption{Study design. A disentanglement dose-response ladder (three checkpoint families, $\lambda_{orth}=0,1,5$, sharing architecture and representation) is used to test whether an implicit assumption---that representation geometry reflects reliability-estimation quality---survives a training intervention that reshapes that geometry. Geometry metrics, three distance-based scorers, and a domain-information probe are computed on the identical embeddings for each of 13 checkpoints (Section~\ref{sec:study-design}).}
\label{fig:schematic}
\end{figure}
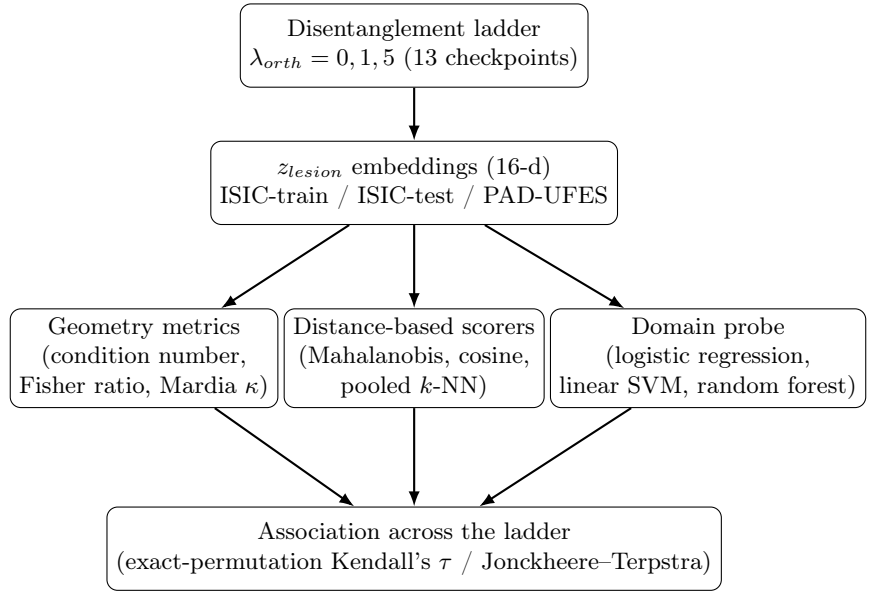

For each of 13 checkpoints (Section~\ref{sec:ladder}), we extract $z_{lesion}$ embeddings for the ISIC training set and for the ISIC-test/PAD-UFES evaluation split. From these embeddings we compute three geometry metrics on the fitted Mahalanobis parameters (Section~\ref{sec:geometry-metrics}), score the evaluation split with eight reliability estimators built on the identical embeddings---five distance-based, three not (Section~\ref{sec:scorers})---and train three supervised probes on the same embeddings to quantify domain-information decodability independent of any of these scorers (Section~\ref{sec:probe}). Association between disentanglement strength, geometry, reliability estimation, and decodability is then tested across the ladder using exact-permutation statistics (Section~\ref{sec:stats}). Figure~\ref{fig:schematic} summarizes this pipeline.

\subsection{Datasets and domains}
\label{sec:datasets}

We use two publicly available skin lesion datasets: ISIC 2018 \citep{codella2019skin,tschandl2018ham10000} as the source domain and PAD-UFES-20 \citep{pacheco2020padufes} as the target domain, following the split and preprocessing pipeline of the underlying disentanglement classifier this study audits. ISIC data is partitioned into training, validation, and test subsets under a fixed, seeded split. PAD-UFES is withheld from label supervision during training. PAD-UFES serves as the adversarial target domain during training and therefore is not independent of the optimization process: the domain-adversarial branch of the audited architecture explicitly optimizes the representation to reduce discriminability between ISIC and PAD-UFES.

\subsection{The disentanglement dose-response ladder}
\label{sec:ladder}

We use three checkpoint families from a domain-adversarial disentanglement architecture, each producing a 16-dimensional lesion representation ($z_{lesion}$). The three families share identical architecture, training recipe, and representation dimensionality, differing only in the strength of an orthogonality regularization penalty ($\lambda_{orth}$) applied between the lesion and context branches (Table~\ref{tab:ladder-design}).

\begin{table}[htbp]
\centering
\caption{The disentanglement dose-response ladder.}
\label{tab:ladder-design}
\begin{tabular}{lccc}
\toprule
Family & $\lambda_{orth}$ & Mechanism & Seeds available \\
\midrule
\texttt{runA\_grl}   & 0 & domain-adversarial component only & 42, 52, 62, 72, 82 (5) \\
\texttt{runB\_orth1} & 1 & + orthogonality term               & 42, 52, 62, 72, 82 (5) \\
\texttt{runB}        & 5 & + stronger orthogonality           & 42, 52, 62 (3) \\
\bottomrule
\end{tabular}
\end{table}

13 (family, seed) checkpoints total. All checkpoints are referenced by explicit, individually verified file path; none is resolved by automated directory search, to avoid a documented failure mode in which the most-recently-modified file in a run directory is not the best-performing validation epoch.

We additionally report a single conventional (non-disentangled) ResNet-50 checkpoint, \texttt{baseline\_soft}, differing from the ladder in architecture, representation dimensionality (2048-d vs.\ 16-d), and training recipe, as a descriptive reference rather than a fourth dose level.

\subsection{Representation geometry metrics}
\label{sec:geometry-metrics}

The selected metrics were pre-specified to represent three complementary aspects of representation geometry---covariance conditioning, class separation, and multivariate normality---rather than to maximize empirical correlation with downstream performance.

\begin{itemize}
\item \textbf{Condition number}: the ratio of the largest to smallest eigenvalue of the regularized, shared within-class covariance matrix used to fit the Mahalanobis estimator (Section~\ref{sec:scorers}), measuring how numerically stable that covariance's inversion is.
\item \textbf{Fisher ratio}: the Hotelling--Lawley trace, $\mathrm{tr}(\Sigma^{-1}S_B)$, computed from the same fitted precision matrix, measuring class separation relative to within-class spread. We additionally report a decoupled scalar companion, $\mathrm{tr}(S_B)/\mathrm{tr}(S_W)$, which involves no matrix inverse and is therefore not entangled with condition number's own estimation noise---the two are reported together because they are not statistically independent measurements when both derive from the same $\Sigma^{-1}$.
\item \textbf{Mardia's multivariate kurtosis}: quantifies departure from the class-conditional Gaussian shape the Mahalanobis estimator assumes. The classical closed-form asymptotic null does not apply once class means and covariance are estimated in-sample from the same residuals being tested; its null distribution is calibrated by parametric bootstrap (200 resamples per checkpoint) rather than the textbook formula.
\end{itemize}

All three metrics are computed on the identical fitted Mahalanobis parameters used for scoring (Section~\ref{sec:scorers}), not from an independently re-estimated covariance, so that geometry and reliability estimation describe the same fitted object.

\subsection{Reliability scorers}
\label{sec:scorers}

Each test embedding is scored by eight structurally distinct rules, applied to identical embeddings and an identical ISIC-train / ISIC-test / PAD-UFES split, for every one of the 13 primary-ladder checkpoints. Five operate only on embedding geometry:

\begin{itemize}
\item \textbf{Mahalanobis distance}: the minimum, over the 8 ISIC classes, of the squared Mahalanobis distance from the test embedding to each class mean, under a shared covariance fit on the ISIC training set with regularization $\varepsilon=10^{-5}$.
\item \textbf{Cosine-to-centroid}: the minimum, over the same 8 class means, of $1-\cos(z,\mu_c)$---identical class-centroid structure to the Mahalanobis scorer, with the covariance/precision step removed, isolating that one variable.
\item \textbf{Pooled $k$-nearest-neighbor distance}: the Euclidean distance from a test embedding to its $k$-th nearest neighbor in the full, class-pooled ISIC training set---no class structure, no covariance, the most assumption-free of the five distance-based scorers \citep{sun2022out}. $k=10$ was specified as the primary/headline value in the analysis plan before any $k$-NN result was computed; $k=1$ and $k=50$ are reported as a prespecified robustness grid, not selected after the fact.
\end{itemize}

Three additional scorers, not distance-based, were evaluated to test whether the failure documented here is specific to distance-based scoring:

\begin{itemize}
\item \textbf{Energy} \citep{liu2020energy}: $-\log\sum_k\exp(\text{logit}_k)$, computed from the classifier head's logits rather than from embedding geometry; a lower log-sum-exp (higher score, in our OOD-is-higher convention) indicates lower classifier confidence.
\item \textbf{ViM (Virtual-Logit Matching)} \citep{wang2022vim}: combines a residual-subspace term---the embedding's norm outside the top principal subspace spanned by the classifier head's weight matrix---with the same logits Energy uses, calibrated on the training set so the two terms are on a comparable scale.
\item \textbf{Density (per-class KDE)}: a Gaussian-kernel density estimate fit separately on each of the 8 ISIC classes' training embeddings (bandwidth by Scott's rule), scored by the negative log-likelihood under the best-fitting class model---replacing Mahalanobis's single-Gaussian assumption with a non-parametric one, while keeping the same per-class-then-best structure.
\end{itemize}

Logits for Energy and ViM were reconstructed from each checkpoint's own classifier head (a linear layer applied directly to the cached $z_{lesion}$ embedding, without a fresh forward pass through the network) and verified to reproduce that checkpoint's own classification accuracy before use. All eight scorers share one score convention (higher score, more out-of-distribution) and are evaluated by AUROC and FPR-at-95\%-TPR on the same ISIC-test-vs-PAD-UFES split. The three non-distance-based scorers were evaluated on the 13 primary-ladder checkpoints only; \texttt{baseline\_soft} (Section~\ref{sec:ladder}) uses a different architecture and classifier head and was not included in this extension.

\subsection{Domain-information probe}
\label{sec:probe}

To test whether domain-discriminative information is present in an embedding independent of whether any of the eight scorers above can access it, we train three supervised classifiers---logistic regression, a linear support vector machine, and a random forest, each at library-default hyperparameters with no tuning---to predict domain membership (ISIC-test vs.\ PAD-UFES) directly from the same $z_{lesion}$ embeddings scored in Section~\ref{sec:scorers}. None of the three probes has access to the original training objective; each is a fresh classifier fit only for this analysis. Probe AUROC is reported as evidence of what is linearly and nonlinearly recoverable from the representation, not as a claim about the training objective's internal computation \citep{hewitt2019designing}.

We report 5-fold stratified cross-validated AUROC on out-of-fold predictions, avoiding the same-data-fit-and-evaluate leakage that would otherwise inflate apparent decodability at this sample size and dimensionality. The ISIC-test/PAD-UFES split is class-imbalanced (approximately 2.2:1); AUROC was selected as the primary probe metric specifically because it is threshold-independent and substantially less sensitive to this imbalance than accuracy-based metrics would be, though it was not supplemented here with an imbalance-corrected metric such as balanced accuracy or precision-recall AUC.

\subsection{Statistical testing}
\label{sec:stats}

The disentanglement ladder has only three ordinal levels with unequal per-level seed counts (5, 5, 3). We test association using Kendall's $\tau$ and, for ordered-group trend testing, the Jonckheere--Terpstra statistic, both evaluated against a full-enumeration exact permutation null---every distinct label arrangement consistent with the true per-rung seed counts---rather than the standard asymptotic approximation, which is not assumed valid at this sample size without checking.

Given the limited number of independent checkpoints ($n=13$), we emphasize consistency across independent analyses rather than statistical significance from any single test; unadjusted exact $p$-values are reported without family-wise correction.

\subsection{Reproducibility}
\label{sec:reproducibility}

Every checkpoint, seed, and preprocessing parameter used in this study is recorded by explicit value or file path; no result in this paper depends on a directory-resolved or otherwise ambiguously-selected checkpoint. Analysis code and derived results supporting this study are available as described in the Data Availability statement accompanying this submission.

\section{Results}
\label{sec:results}

\subsection{Representation geometry changes with disentanglement strength}
\label{sec:results-geometry}

\begin{figure}[htbp]
\centering
\includegraphics[width=\textwidth]{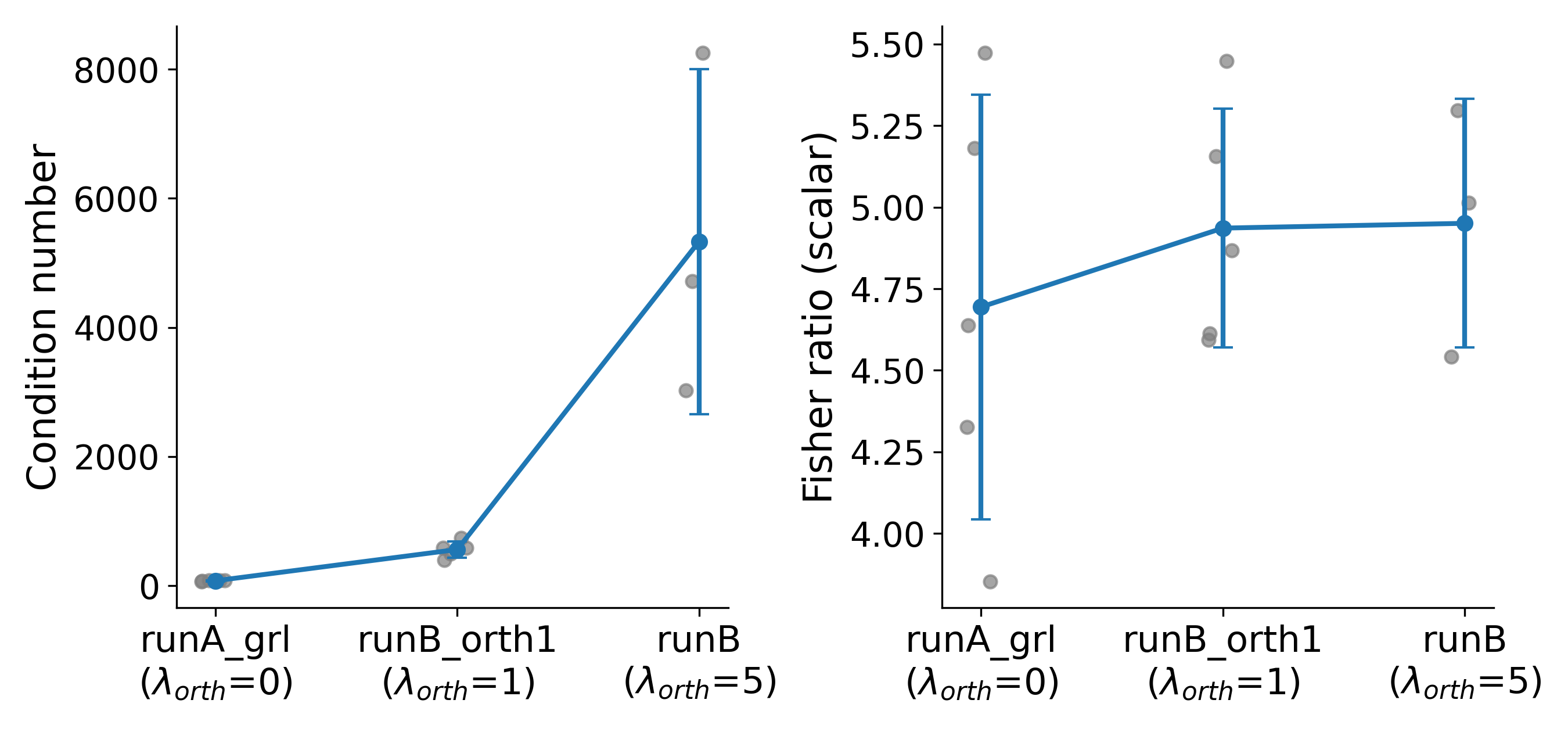}
\caption{Representation geometry across the disentanglement ladder. Condition number (left) and the decoupled Fisher-ratio scalar $\mathrm{tr}(S_B)/\mathrm{tr}(S_W)$ (right) for each of 13 checkpoints (gray points; mean $\pm$ SD in blue), across three levels of orthogonality strength. Condition number increases by approximately two orders of magnitude with $\lambda_{orth}$ (Kendall's $\tau=0.84$, exact $p=2.8\times10^{-5}$; Table~\ref{tab:geometry-summary}); the Fisher-ratio scalar shows no significant trend. The remaining three geometry metrics (Fisher ratio, Hotelling--Lawley form; Mardia's kurtosis $b$ and $z$) are in Supplementary Figure~S1.}
\label{fig:ladder-trend}
\end{figure}

Condition number increased monotonically with $\lambda_{orth}$: $75.5\pm7.8$ (\texttt{runA\_grl}), $559.8\pm125.9$ (\texttt{runB\_orth1}), $5329.9\pm2672.2$ (\texttt{runB})---a two-order-of-magnitude change across the ladder (Table~\ref{tab:geometry-summary}, Figure~\ref{fig:ladder-trend}). This trend was significant by exact-permutation Kendall's $\tau$ ($\tau=0.84$, $p=2.8\times10^{-5}$, $n=13$) and sign-stable on the common-seed subset ($\tau=0.87$, $p=0.0012$, $n=9$), and confirmed by Jonckheere--Terpstra trend testing ($J=55.0$ against a null mean of 27.5, exact $p=1.4\times10^{-5}$).

The remaining four geometry metrics showed no significant trend: Fisher ratio (Hotelling--Lawley), $\tau=0.17$, $p=0.52$; Fisher ratio (decoupled scalar), $\tau=0.08$, $p=0.80$; Mardia's kurtosis ($b$), $\tau=-0.14$, $p=0.61$; Mardia's kurtosis ($z$), $\tau=-0.17$, $p=0.52$ (Table~\ref{tab:geometry-summary}).

Increasing $\lambda_{orth}$ is associated with a large, statistically significant change in condition number; the other four geometry metrics tested show no significant association with $\lambda_{orth}$.

\begin{table}[htbp]
\centering
\caption{Representation geometry across the disentanglement ladder (mean $\pm$ SD).}
\label{tab:geometry-summary}
\begin{tabular}{lccc}
\toprule
Metric & \texttt{runA\_grl} ($\lambda_{orth}=0$) & \texttt{runB\_orth1} ($\lambda_{orth}=1$) & \texttt{runB} ($\lambda_{orth}=5$) \\
\midrule
Condition number                          & $75.5 \pm 7.8$              & $559.8 \pm 125.9$           & $5329.9 \pm 2672.2$ \\
Fisher ratio (Hotelling--Lawley, $\times10^{5}$) & $5.41 \pm 0.81$      & $5.71 \pm 0.43$              & $5.63 \pm 0.50$ \\
Fisher ratio (decoupled scalar)           & $4.69 \pm 0.65$              & $4.94 \pm 0.37$              & $4.95 \pm 0.38$ \\
Mardia's kurtosis ($b$)                   & $739.9 \pm 86.5$             & $788.0 \pm 240.9$            & $679.9 \pm 63.7$ \\
Mardia's kurtosis ($z$)                   & $341.1 \pm 99.3$             & $397.3 \pm 257.6$            & $270.7 \pm 53.4$ \\
\midrule
$n$ seeds                                 & 5                            & 5                            & 3 \\
\bottomrule
\end{tabular}
\end{table}

\subsection{Representation geometry shows no measurable association with Mahalanobis AUROC}
\label{sec:results-geometry-auroc}

Mahalanobis AUROC (ISIC-test vs.\ PAD-UFES) was $0.401\pm0.028$ (\texttt{runA\_grl}), $0.408\pm0.022$ (\texttt{runB\_orth1}), and $0.393\pm0.025$ (\texttt{runB})---below the chance value of 0.5 at every rung (Table~\ref{tab:mahalanobis-auroc}).

\begin{figure}[htbp]
\centering
\includegraphics[width=0.6\textwidth]{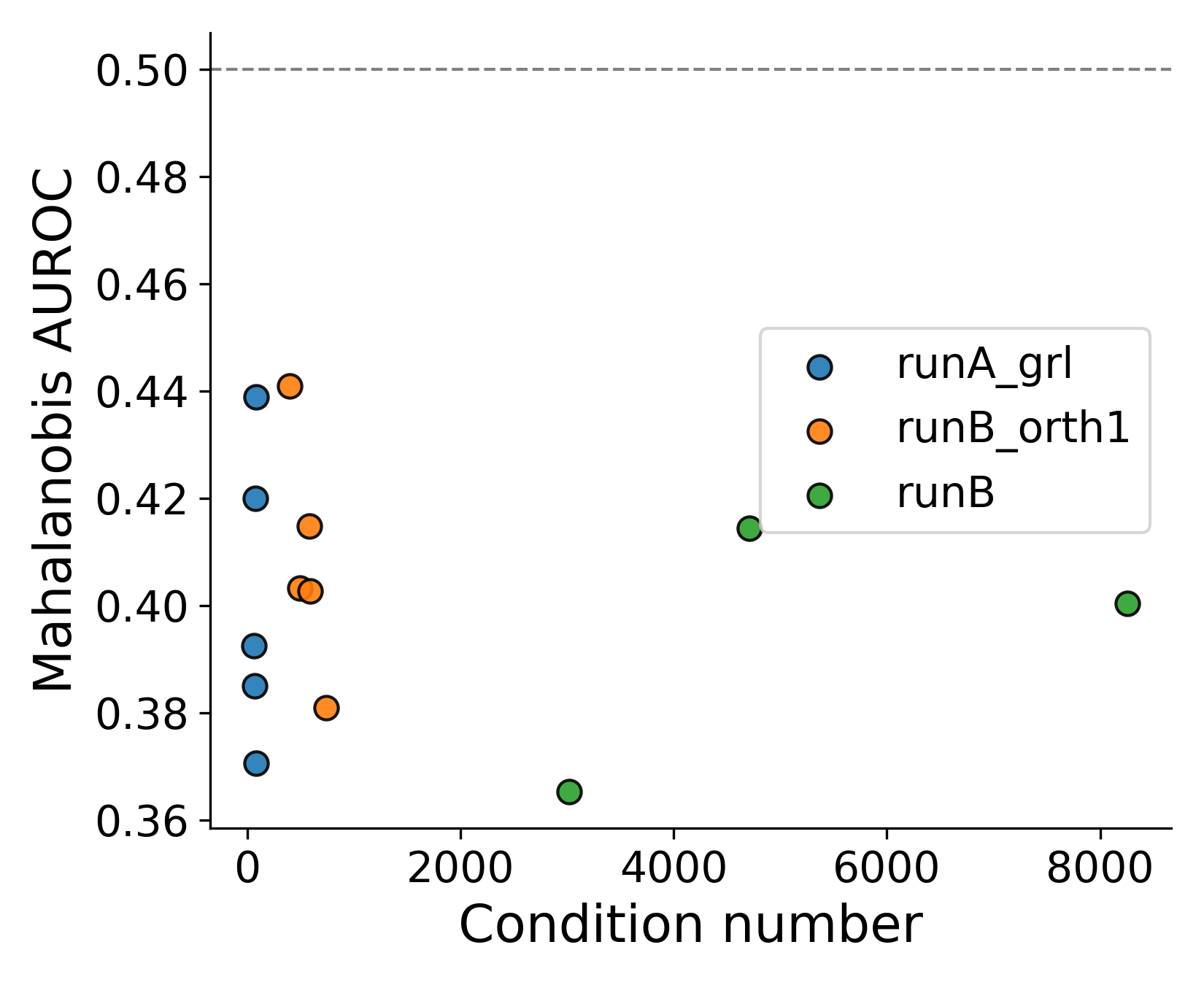}
\caption{Representation geometry does not predict Mahalanobis AUROC. Condition number vs.\ Mahalanobis-distance AUROC (ISIC-test vs.\ PAD-UFES), one point per checkpoint, colored by rung. No association is apparent (Kendall's $\tau=-0.13$, $p=0.59$; Table~\ref{tab:mahalanobis-auroc}); the remaining four geometry metrics show the same pattern (Supplementary Figure~S2).}
\label{fig:geometry-vs-auroc}
\end{figure}

None of the five geometry metrics, including condition number, showed a significant association with Mahalanobis AUROC: condition number, $\tau=-0.13$, $p=0.59$; Fisher ratio (HL), $\tau=-0.28$, $p=0.20$; Fisher ratio (scalar), $\tau=-0.15$, $p=0.51$; Mardia's kurtosis ($b$), $\tau=-0.13$, $p=0.59$; Mardia's kurtosis ($z$), $\tau=-0.13$, $p=0.59$ (Figure~\ref{fig:geometry-vs-auroc}). AUROC itself showed no significant trend across the ladder (Jonckheere--Terpstra, $J=25.0$ against a null mean of 27.5, exact $p=0.65$).

An approximately two-orders-of-magnitude change in condition number (Section~\ref{sec:results-geometry}) is not accompanied by a significant change in Mahalanobis AUROC; no geometry metric tested shows a significant association with AUROC.

\begin{table}[htbp]
\centering
\caption{Mahalanobis-distance AUROC (ISIC-test vs.\ PAD-UFES), by rung (mean $\pm$ SD across seeds).}
\label{tab:mahalanobis-auroc}
\begin{tabular}{lcccc}
\toprule
Rung & $\lambda_{orth}$ & $n$ seeds & AUROC & FPR@95\%TPR \\
\midrule
\texttt{runA\_grl}   & 0 & 5 & $0.401 \pm 0.028$ & $0.985 \pm 0.009$ \\
\texttt{runB\_orth1} & 1 & 5 & $0.408 \pm 0.022$ & $0.980 \pm 0.012$ \\
\texttt{runB}        & 5 & 3 & $0.393 \pm 0.025$ & $0.985 \pm 0.003$ \\
\bottomrule
\end{tabular}
\end{table}

\subsection{Distance-based scores are systematically lower for the target domain}
\label{sec:results-distance-gap}

Median Mahalanobis squared distance was lower for PAD-UFES than for ISIC-test at every one of the 13 checkpoints. Pooled across seeds within each rung: \texttt{runA\_grl}, ISIC-test median 13.50 vs.\ PAD-UFES median 8.57 ($n_{id}=25335$, $n_{ood}=11490$); \texttt{runB\_orth1}, 14.10 vs.\ 8.99 ($n_{id}=25335$, $n_{ood}=11490$); \texttt{runB}, 13.20 vs.\ 8.08 ($n_{id}=15201$, $n_{ood}=6894$). Median distance was lower for the target domain than the source domain at all 13 checkpoints.

Feature-norm distributions differed modestly between domains. Pooled median $\|z\|$: \texttt{runA\_grl}, ISIC-test 6.31 vs.\ PAD-UFES 6.21; \texttt{runB\_orth1}, 6.38 vs.\ 6.28; \texttt{runB}, 6.28 vs.\ 5.97. At the level of pooled means, the direction was not consistent across rungs (\texttt{runA\_grl} mean: 6.72 ISIC-test vs.\ 6.83 PAD-UFES). The direction of this difference was not consistent at the level of means.

Nearest-centroid classification assigned PAD-UFES samples to the majority ISIC class (Nevus) at 61.6\% (\texttt{runA\_grl}), 61.9\% (\texttt{runB\_orth1}), and 63.9\% (\texttt{runB}), compared to 53.0\%, 53.1\%, and 53.4\% respectively for ISIC-test samples classified against their own nearest centroid. This was 8.7--10.5 percentage points higher than the ISIC-test baseline rate (ratio 1.16--1.20).

\begin{figure}[htbp]
\centering
\includegraphics[width=\textwidth]{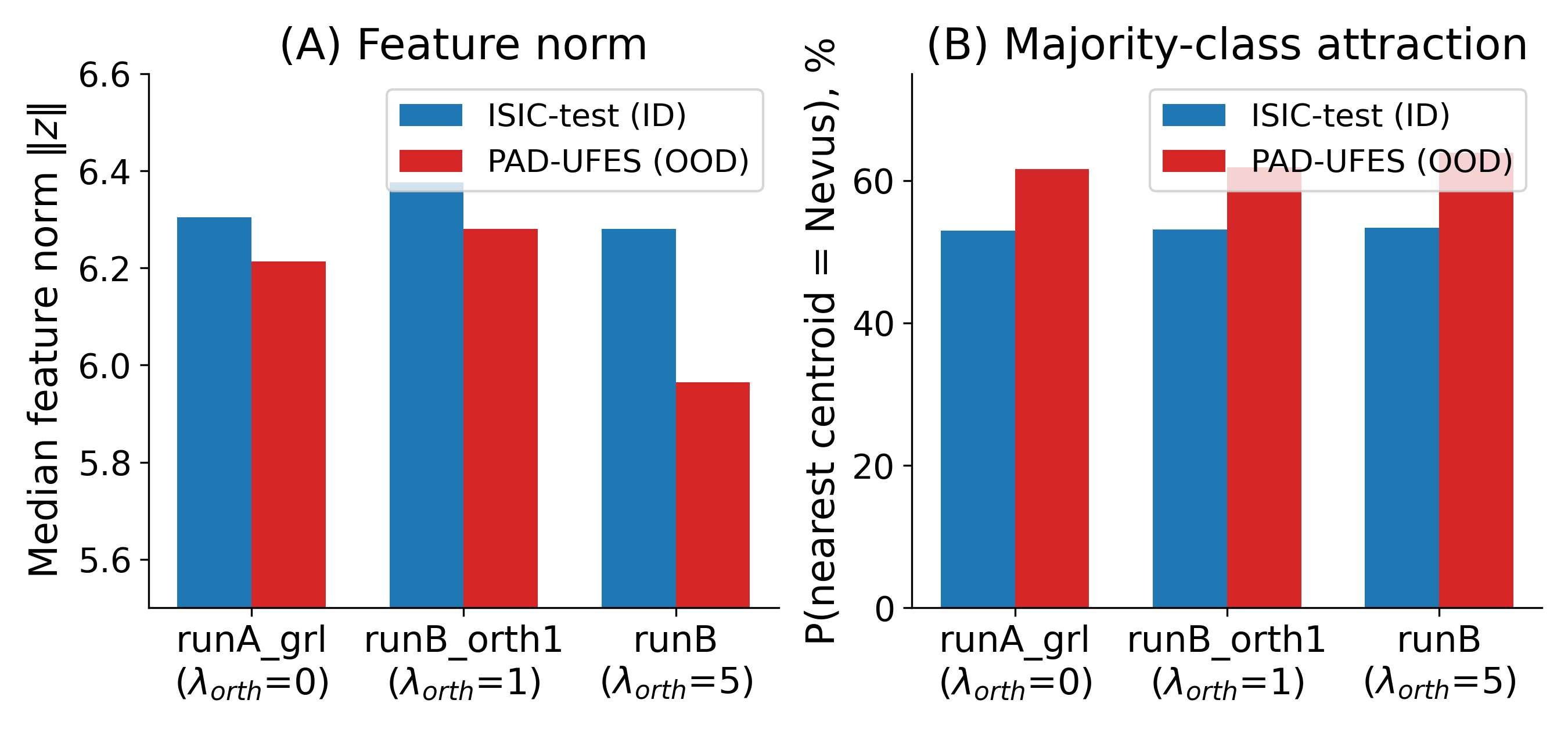}
\caption{Neither feature-norm difference nor majority-class attraction fully accounts for the raw ID/OOD distance gap. (A) Pooled median feature norm, ISIC-test vs.\ PAD-UFES, per rung. (B) Nearest-centroid assignment of ISIC-test and PAD-UFES samples to the majority ISIC class (Nevus), per rung. The norm difference is modest and not consistent in direction at the level of means; PAD-UFES's excess attraction to the majority class (8.7--10.5 percentage points above the ISIC-test baseline) is real but does not, on its own, account for the full distance separation in Figure~\ref{fig:scorer-comparison}.}
\label{fig:mechanism}
\end{figure}

\subsection{The failure generalizes across eight scorer families}
\label{sec:results-scorer-comparison}

\begin{figure}[htbp]
\centering
\includegraphics[width=\textwidth]{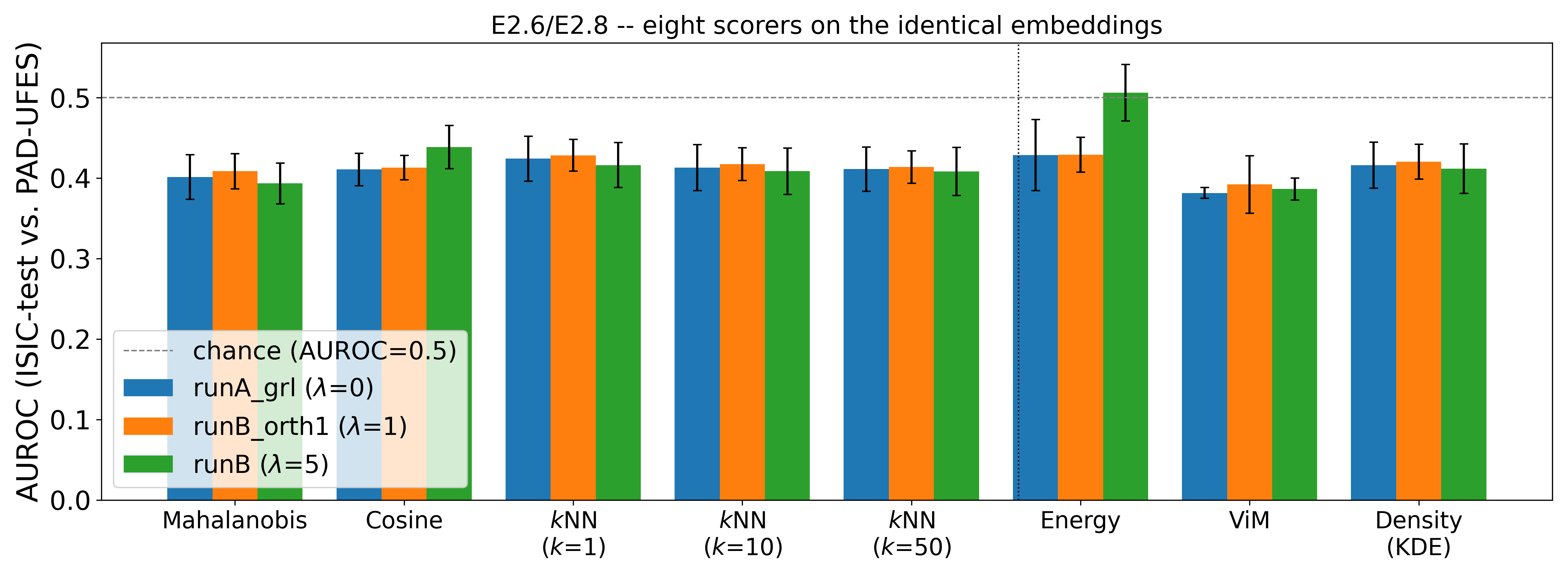}
\caption{Eight scorers fail alike, with one isolated exception. AUROC (ISIC-test vs.\ PAD-UFES) for five distance-based scorers (Mahalanobis distance, cosine-to-centroid similarity, pooled $k$-nearest-neighbor distance at $k=1,10,50$) and three non-distance-based scorers (Energy, ViM, per-class KDE density), computed on the identical embeddings, grouped by rung (mean $\pm$ SD across seeds). Dashed line: chance (AUROC = 0.5); dotted line separates the two scorer families. Seven of eight scorers fall within a 0.39--0.42 range with no significant trend; Energy alone rises with $\lambda_{orth}$ (Jonckheere--Terpstra exact $p=0.032$), discussed in Section~\ref{sec:discussion} as an isolated, hypothesis-generating observation (Table~\ref{tab:scorer-summary}).}
\label{fig:scorer-comparison}
\end{figure}

Pooled across all 13 checkpoints, AUROC was $0.402\pm0.024$ (Mahalanobis), $0.418\pm0.022$ (cosine-to-centroid), $0.424\pm0.023$ ($k$-NN, $k=1$), $0.414\pm0.024$ ($k$-NN, $k=10$), and $0.411\pm0.023$ ($k$-NN, $k=50$) (Table~\ref{tab:scorer-summary}, Figure~\ref{fig:scorer-comparison}). All five values fall within a 0.40--0.42 range and below the chance value of 0.5. Despite relying on substantially different scoring formulations---parametric with covariance, parametric without covariance, and non-parametric---all five distance-based scorer families produced highly similar AUROC values.

To test whether this failure is specific to distance-based scoring, we additionally evaluated three scorers that are not distance-based: Energy (a logit-based confidence score), ViM (Virtual-Logit Matching, a residual-subspace scorer that combines geometry and logits), and a per-class kernel density estimator. Pooled across the same 13 checkpoints, AUROC was $0.447\pm0.047$ (Energy), $0.387\pm0.022$ (ViM), and $0.417\pm0.025$ (density estimator)---all three within the same 0.39--0.45 range as the five distance-based scorers, and below or near the chance value of 0.5.

None of the five distance-based scorer variants showed a significant association with $\lambda_{orth}$: Mahalanobis, $\tau=-0.08$, $p=0.80$; cosine, $\tau=0.32$, $p=0.19$; $k$-NN ($k=1$), $\tau=-0.05$, $p=0.90$; $k$-NN ($k=10$), $\tau=0.02$, $p=1.00$; $k$-NN ($k=50$), $\tau=0.08$, $p=0.80$. Jonckheere--Terpstra trend testing gave the same result for all five (exact $p$ between 0.10 and 0.65). Among the three additional scorers, ViM ($\tau=0.08$, $p=0.80$) and the density estimator ($\tau=0.02$, $p=1.00$) likewise showed no significant association. Energy exhibited a modest monotonic trend with increasing disentanglement (Kendall's $\tau=0.44$, exact $p=0.065$; Jonckheere--Terpstra $p=0.032$), whereas the remaining seven scorers showed no consistent trend (full statistics in Supplementary Table~S7).

Eight structurally distinct scorers---five distance-based (parametric with and without covariance, and non-parametric $k$-NN at three values of $k$) and three that are not distance-based (a logit-based confidence score, a residual-subspace scorer, and a non-parametric density estimator)---report AUROC in essentially the same 0.39--0.45 range, all at or below chance, with one exception (Energy's Jonckheere--Terpstra trend) discussed in Section~\ref{sec:discussion}.

\begin{table}[htbp]
\centering
\caption{AUROC (ISIC-test vs.\ PAD-UFES) for eight scorers, by rung and pooled (mean $\pm$ SD across seeds).}
\label{tab:scorer-summary}
\begin{tabular}{lcccc}
\toprule
Scorer & \texttt{runA\_grl} & \texttt{runB\_orth1} & \texttt{runB} & Pooled ($n=13$) \\
\midrule
\multicolumn{5}{l}{\textit{Distance-based}} \\
Mahalanobis         & $0.401 \pm 0.028$ & $0.408 \pm 0.022$ & $0.393 \pm 0.025$ & $0.402 \pm 0.024$ \\
Cosine-to-centroid  & $0.411 \pm 0.020$ & $0.413 \pm 0.015$ & $0.439 \pm 0.027$ & $0.418 \pm 0.022$ \\
$k$-NN ($k=1$)      & $0.424 \pm 0.028$ & $0.428 \pm 0.020$ & $0.416 \pm 0.028$ & $0.424 \pm 0.023$ \\
$k$-NN ($k=10$)     & $0.413 \pm 0.029$ & $0.417 \pm 0.020$ & $0.408 \pm 0.029$ & $0.414 \pm 0.024$ \\
$k$-NN ($k=50$)     & $0.411 \pm 0.027$ & $0.414 \pm 0.020$ & $0.408 \pm 0.030$ & $0.411 \pm 0.023$ \\
\midrule
\multicolumn{5}{l}{\textit{Non-distance-based}} \\
Energy              & $0.429 \pm 0.044$ & $0.429 \pm 0.022$ & $0.506 \pm 0.035$ & $0.447 \pm 0.047$ \\
ViM                 & $0.381 \pm 0.007$ & $0.392 \pm 0.036$ & $0.386 \pm 0.014$ & $0.387 \pm 0.022$ \\
Density (KDE)       & $0.416 \pm 0.029$ & $0.420 \pm 0.022$ & $0.412 \pm 0.031$ & $0.417 \pm 0.025$ \\
\bottomrule
\end{tabular}
\end{table}

\subsection{Domain information remains decodable despite distance-based inaccessibility}
\label{sec:results-probe}

\begin{figure}[htbp]
\centering
\includegraphics[width=0.8\textwidth]{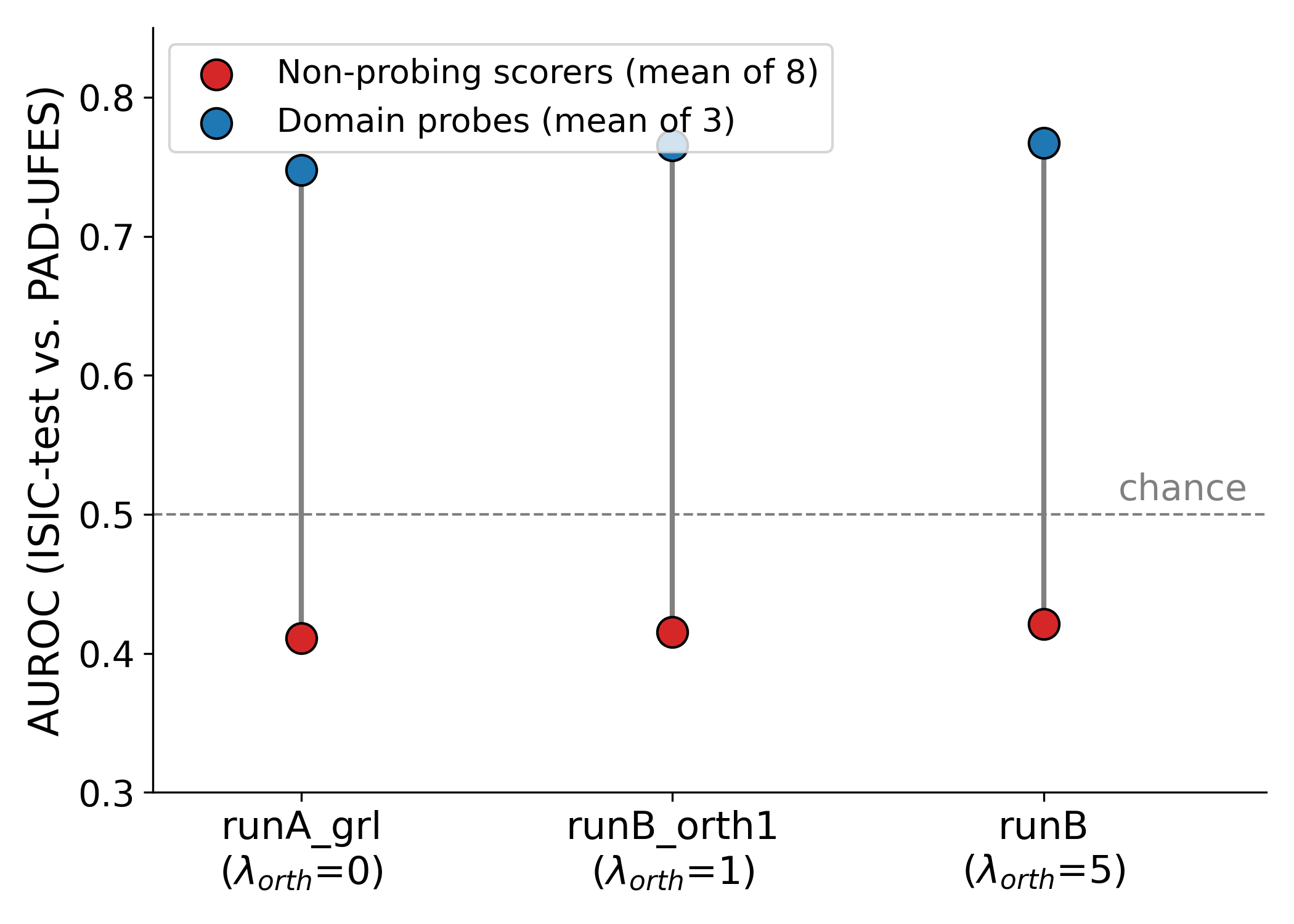}
\caption{Domain information is decodable but not accessible to non-probing scorers. For each rung, mean AUROC across the eight scorers in Figure~\ref{fig:scorer-comparison} (five distance-based, three non-distance-based) is connected to mean AUROC across three independent domain probes---logistic regression, linear SVM, random forest (Section~\ref{sec:probe})---computed on the identical embeddings. Non-probing scorers cluster at or below chance ($\approx 0.41$); probes recover domain membership at 0.72--0.81 AUROC at every $\lambda_{orth}$ level. \texttt{baseline\_soft} (Section~\ref{sec:results-baseline}) is not plotted on this axis, as it is a descriptive reference confounded by architecture and dimensionality (Section~\ref{sec:ladder}), not a controlled comparison point.}
\label{fig:dumbbell}
\end{figure}

On the identical embeddings scored in Section~\ref{sec:results-scorer-comparison}, three supervised probes recovered domain membership (ISIC-test vs.\ PAD-UFES) at: \texttt{runA\_grl}, 0.719 (logistic regression), 0.717 (linear SVM), 0.807 (random forest); \texttt{runB\_orth1}, 0.741, 0.740, 0.815; \texttt{runB}, 0.750, 0.751, 0.800 (Table~\ref{tab:probe-vs-distance}, Figure~\ref{fig:dumbbell}). AUROC exceeded 0.70 for all three probe families at every rung.

None of the three probes showed a significant association with $\lambda_{orth}$: logistic regression, $\tau=0.26$, $p=0.30$; linear SVM, $\tau=0.26$, $p=0.30$; random forest, $\tau=-0.11$, $p=0.70$.

Three independent classifiers, applied to the same embeddings scored by the eight non-probing scorers in Section~\ref{sec:results-scorer-comparison}, recover domain membership at 0.72--0.81 AUROC at every $\lambda_{orth}$ level tested; probe AUROC shows no significant association with $\lambda_{orth}$.

%
\begin{table}[htbp]
\centering
\caption{Non-probing scorer AUROC (pooled across Table~\ref{tab:scorer-summary}) vs.\ domain-probe AUROC, by rung. \texttt{baseline\_soft} is a descriptive reference, not a fourth ladder point (Section~\ref{sec:ladder}).}
\label{tab:probe-vs-distance}
\resizebox{\textwidth}{!}{%
\begin{tabular}{llcccc}
\toprule
Rung & Role & Scorers (pooled) & Probe (LR) & Probe (SVM) & Probe (RF) \\
\midrule
\texttt{runA\_grl}   & primary ladder & $0.411$ & $0.719$ & $0.717$ & $0.807$ \\
\texttt{runB\_orth1} & primary ladder & $0.415$ & $0.741$ & $0.740$ & $0.815$ \\
\texttt{runB}        & primary ladder & $0.421$ & $0.750$ & $0.751$ & $0.800$ \\
\midrule
\texttt{baseline\_soft} & categorical reference ($n=1$) & $0.828$ & $0.998$ & $0.997$ & $0.977$ \\
\bottomrule
\end{tabular}%
}
\end{table}

\subsection{Descriptive reference: a non-disentangled baseline}
\label{sec:results-baseline}

For a single ResNet-50 checkpoint (\texttt{baseline\_soft}; one seed, architecture and representation dimensionality differing from the ladder, Section~\ref{sec:ladder}), pooled AUROC across the five distance-based scorers was 0.828 (the three non-distance-based scorers were not evaluated on this architecture, Section~\ref{sec:results-scorer-comparison}), and probe AUROC was 0.998 (logistic regression), 0.997 (linear SVM), and 0.977 (random forest) (Table~\ref{tab:probe-vs-distance}).

\section{Discussion}
\label{sec:discussion}

\subsection{Finding}
\label{sec:discussion-finding}

Increasing disentanglement strength substantially changed representation geometry without improving distance-based reliability estimation, while the domain information such estimation would require remained recoverable from the same representations by a non-distance-based reader.

\subsection{Interpretation}
\label{sec:discussion-interpretation}

This study audited a simple assumption: changing representation geometry should change the behavior of distance-based reliability estimators. The first part of this assumption held---condition number moved by two orders of magnitude across the disentanglement ladder (Section~\ref{sec:results-geometry}). The second did not: no geometry metric tracked Mahalanobis AUROC, and AUROC itself showed no trend across the ladder (Section~\ref{sec:results-geometry-auroc}).

A null result admits two straightforward explanations: that Mahalanobis distance specifically is a poor instrument, or that disentanglement training erased the domain-discriminative information a reliability estimator would need. Neither holds. Cosine-to-centroid similarity discards the covariance structure Mahalanobis distance depends on; pooled $k$-nearest-neighbor distance discards the notion of a class centroid altogether; Energy \citep{liu2020energy} discards representation geometry altogether in favor of the classifier's own logits; ViM \citep{wang2022vim} combines a residual-subspace geometric term with those same logits; a per-class kernel density estimator replaces Mahalanobis's single-Gaussian assumption with a non-parametric one. Despite these substantially different scoring principles, seven of the eight scorers tested converged on the same 0.39--0.42 AUROC (Section~\ref{sec:results-scorer-comparison})---a failure shared across formulas this different reflects what they share, not a defect specific to any one. Nor was the relevant information absent: three probes with no access to the training objective recovered domain membership at 0.72--0.81 AUROC at every disentanglement level tested (Section~\ref{sec:results-probe}), well above what these scoring rules reported on the identical embeddings. The information was not gone. It was present, recoverable, and specifically unreachable by seven of the eight scoring rules tested.

The eighth scorer, Energy, is an exception we report plainly rather than explain away. Among the eight evaluated scorers, Energy exhibited a modest monotonic trend with increasing disentanglement (Jonckheere--Terpstra exact $p=0.032$), whereas the remaining seven scorers showed no consistent trend (Section~\ref{sec:results-scorer-comparison}). This isolated observation should be interpreted cautiously: it is one result out of eight scorers tested without correction for multiple comparisons, it was not pre-registered, its Kendall's $\tau$ (0.44, exact $p=0.065$) does not clear the $|\tau|\geq0.3$-and-significant threshold this study otherwise applies (Section~\ref{sec:results-geometry-auroc}), and the effect size is modest. We treat it as a hypothesis-generating observation rather than as evidence against the main conclusion, which we accordingly scope to seven of the eight scorers tested rather than all eight. That Energy alone showed this trend suggests that accessibility may depend not only on representation geometry but also on the inductive biases of individual scoring functions---whether this reflects a genuine property of energy-based scoring or a dataset-specific effect is an open question, as is whether it persists under different pretraining paradigms or model architectures.

Taken together, these findings indicate that \textbf{information can remain decodable while becoming largely inaccessible to non-probing reliability estimators.} This is not a claim that any one of these methods is poorly designed, and not a claim that disentanglement training destroys information. It is a claim about a gap between two properties of a representation---whether information is present, and whether it is organized in a form a specific family of estimators can use---that the ladder design in this study makes it possible to separate for the first time in this setting. This claim is scoped to the training regime studied here---domain-adversarial disentanglement evaluated against its own adversarial target domain (Section~\ref{sec:discussion-limitations})---not to representation learning or domain shift in general.

Section~\ref{sec:results-distance-gap}'s supporting evidence (a modest, direction-inconsistent norm difference; a majority-class attraction of 8.7--10.5 percentage points above baseline) indicates that the raw distance gap between domains is real but is not fully accounted for by either mechanism individually; neither is proposed here as a complete account of \emph{why} the gap exists. Understanding the origin of this systematic below-chance behavior remains future work.

\subsection{Relation to literature}
\label{sec:discussion-literature}

Mahalanobis-distance OOD detection \citep{lee2018simple} and its non-parametric successors, including pooled $k$-nearest-neighbor scoring \citep{sun2022out}, are typically evaluated by benchmarking AUROC across datasets and architectures, not by auditing whether the geometric assumption connecting representation and reliability estimate survives a specific training intervention. Energy-based scoring \citep{liu2020energy} and Virtual-Logit Matching \citep{wang2022vim} were developed specifically to move beyond a purely geometric notion of reliability by incorporating the classifier's own logits, making them a natural extension of the audit rather than a different question. This study does not propose a competitor to any of these methods; all five distance-based and three non-distance-based scorers are used here as instruments for the audit, alongside a cosine-similarity variant and a kernel density estimator that isolate specific structural assumptions from one another. The finding that seven of the eight degrade identically under this training regime is a statement about a premise most of these methods share, not a comparative ranking of them.

Domain-adversarial and orthogonality-based disentanglement training \citep{ganin2015unsupervised} is one of several strategies proposed to reduce reliance on shortcut features in deep classifiers \citep{geirhos2020shortcut}. That literature evaluates disentanglement primarily by its effect on classification robustness and shortcut reliance; this study evaluates a different, downstream consequence---what disentanglement does to a reliability estimator built on top of the resulting representation---and finds that the two do not move together. This does not bear on whether disentanglement succeeds at its stated goal, only on whether a distance-based reliability estimator layered on top of it continues to function as it did before the intervention.

The distinction this study draws between the presence of information in a representation and a specific estimator's ability to use it is closely related to, and constrained by, the probing-classifier literature's own caution that a supervised probe's success does not establish that a model itself uses the decoded information in the way the probe accesses it \citep{hewitt2019designing}. This study makes a narrower claim than that literature warns against overreaching toward: the probes here are not offered as evidence about what the disentanglement objective computes internally, only as an independent measurement of what remains extractable from its output representation, without characterizing the mechanism by which that representation was produced.

More broadly, these findings suggest that evaluating learned representations only through downstream predictive performance may overlook an equally important question: whether the representation is organized in a form that downstream reliability estimators can actually exploit. Geometry, decodability, and distance accessibility therefore appear to be related but distinct properties, and improvement in one should not automatically be interpreted as improvement in the others.

For practitioners combining domain-adversarial disentanglement training with a distance-based reliability gate in a deployed pipeline, this audit implies a specific caution: such a gate may not reliably flag inputs from the domain the disentanglement objective was trained against, even where it performs as expected against unrelated distributions. Where feasible, reliability estimation in such pipelines should be validated directly against the domain the training procedure targets, rather than assumed to transfer from general model qualification.

\subsection{Limitations}
\label{sec:discussion-limitations}

PAD-UFES is not an arbitrary target domain in this design: it is the domain the disentanglement architecture's adversarial branch is trained against. Since PAD-UFES also serves as the adversarial domain during training, our conclusions should be interpreted as characterizing representations learned under this training regime rather than all domain shifts.

The comparison against a conventional, non-disentangled baseline (Section~\ref{sec:results-baseline}) differs from the disentanglement ladder in architecture, representation dimensionality, and training recipe simultaneously, and is drawn from a single checkpoint. It is reported descriptively and does not support any causal attribution of the ladder's findings to disentanglement specifically, as opposed to architecture or dimensionality; a matched, non-adversarial checkpoint at the same architecture and 16-dimensional representation would be needed to make that attribution, and no such checkpoint exists in the inventory available for this study. Attributing the central finding specifically to disentanglement, rather than to architecture or dimensionality, remains an important direction.

The ladder itself spans three ordinal treatment levels with unequal per-level seed counts ($n=13$ total, $n=9$ on the common-seed subset), which limits the statistical power of any null result reported here---the absence of a significant association is evidence of a bounded effect size at this sample size, not evidence that no association exists at any sample size. Exact-permutation testing was used throughout in place of asymptotic approximations not verified to hold at this scale, and $p$-values are reported without family-wise correction across the multiple metrics and scorers tested, on the stated basis that conclusions rest on consistency of effect across independent analyses rather than on any single threshold crossing; a reader applying a stricter correction should note that the one large, load-bearing effect in this study (condition number's association with $\lambda_{orth}$) survives any standard correction, while every null result was already reported as non-significant before any correction was considered. The null associations reported in Sections~\ref{sec:results-geometry-auroc}--\ref{sec:results-probe} should be read as bounded by the modest power available at $n=13$, not as evidence that no association exists at any sample size.

This study does not include a replication on a domain independent of the ISIC/PAD-UFES pair, nor a check of generalization beyond the single architecture audited here. Both were considered and are not included: a third-domain replication would require a dataset without documented image overlap with the training data, which was not available within the scope of this study, and a cross-architecture check was not pursued for the same reason. Neither omission is filled by the baseline reference described above, which addresses architecture and dimensionality but not domain independence. The bootstrap check proposed elsewhere to quantify shared estimation noise between condition number and Fisher ratio, both derived from the same fitted precision matrix, was not performed; condition number, not Fisher ratio, is the metric driving the geometric finding in Section~\ref{sec:results-geometry}, which bounds but does not eliminate this as a source of uncertainty in the broader geometry measurement. Checkpoint selection was verified by explicit file path rather than by directory search, but was not independently cross-checked against logged validation curves beyond the single file available per run. Mardia's kurtosis null distribution was calibrated with 200 bootstrap resamples per checkpoint (Section~\ref{sec:geometry-metrics}); the Mahalanobis regularization constant ($\varepsilon=10^{-5}$) and the 16-dimensional representation itself were inherited from the audited classifier rather than varied as part of this study. ``Reliability estimation'' throughout this study refers specifically to out-of-distribution detection AUROC; broader constructs such as calibration or selective prediction are not addressed.

This distinction is particularly relevant when domain-adversarial representations are combined with distance-based or logit-based uncertainty estimation in safety-critical medical AI systems. The same information can remain decodable while becoming largely inaccessible to non-probing reliability estimators.

\section{Conclusion}
\label{sec:conclusion}

This study audited whether disentanglement training that reshapes representation geometry preserves the assumption distance-based reliability estimation depends on. It largely does not: representation geometry changed substantially, and reliability estimation largely did not track it---a failure shared by seven of eight structurally distinct scoring rules tested, including three that are not distance-based. The eighth, an energy-based scorer, showed an isolated trend that we treat as hypothesis-generating rather than as evidence against this pattern (Section~\ref{sec:discussion}). The information these scorers would have needed remained recoverable from the same representations by a supervised probe, indicating that the two properties---whether information is present in a representation and whether it is organized in a form a given estimator can use---can diverge under this training regime. Evaluating learned representations for downstream reliability estimation may therefore require assessing both properties directly, rather than assuming one implies the other. Whether this gap persists under different pretraining paradigms or model architectures remains an open question for future work.

\section*{Acknowledgements}
The author gratefully acknowledges the ISIC Archive and HAM10000 contributors for making the ISIC 2018 dataset publicly available, and the PAD-UFES-20 team for providing the clinical photography dataset. Computational resources were provided by Hanoi University of Science and Technology.

\section*{Declarations}

\subsection*{Ethical approval}
This study used publicly available de-identified datasets; therefore, ethical approval was not required.

\subsection*{Funding}
This research received no external funding.

\subsection*{Competing interests}
The author declares no competing interests.

\subsection*{CRediT authorship contribution statement}
D.-V. Tran contributed to conceptualization, methodology, software, validation, formal analysis, investigation, resources, data curation, visualization, writing---original draft, and writing---review and editing.

\subsection*{Data availability}
Both datasets used in this study are publicly available. The ISIC 2018 dataset is accessible through the ISIC Archive (\url{https://challenge.isic-archive.com/data/}). The PAD-UFES-20 dataset is available at \url{https://data.mendeley.com/datasets/zr7vgbcyr2}. Code will be made publicly available upon acceptance.

\bibliographystyle{elsarticle-num}
\bibliography{refs}

\end{document}


\maketitle
\vspace{-1.5em}

\section*{Supplementary Material}

\subsection*{S1. Full geometry metrics across the ladder}

\begin{figure}[htbp]
\centering
\includegraphics[width=\textwidth]{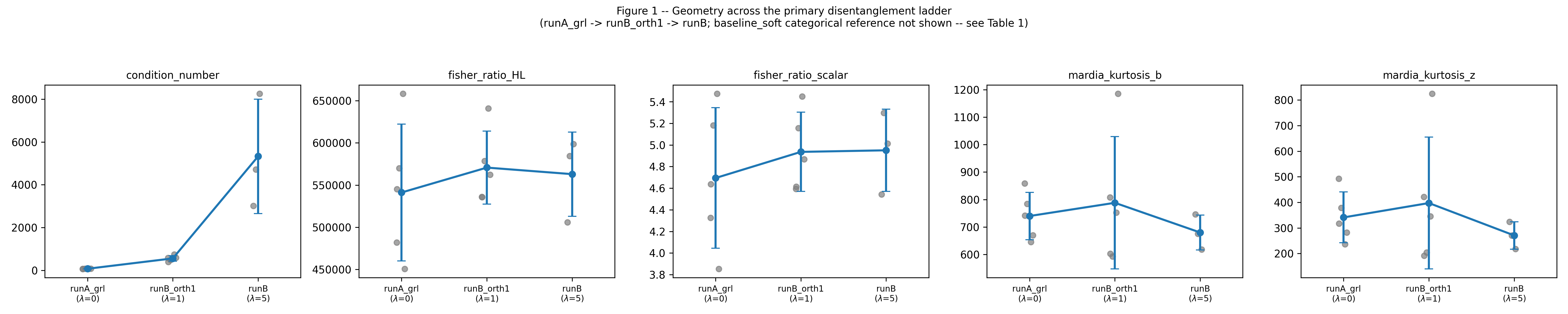}
\caption{All five geometry metrics across the disentanglement ladder.}
\label{fig:s1}
\end{figure}

Extends main-text Figure~\ref{fig:ladder-trend}, which shows only condition number and the decoupled Fisher-ratio scalar. Figure~\ref{fig:s1} shows all five pre-specified geometry metrics (condition number, Fisher ratio---Hotelling--Lawley and decoupled-scalar forms---and Mardia's multivariate kurtosis, $b$ and $z$ statistics), one gray point per checkpoint ($n=13$), mean $\pm$ SD in blue, across the three rungs of the disentanglement ladder. Table~\ref{tab:s1-geometry-stats} reports the corresponding exact-permutation Kendall's $\tau$ and Jonckheere--Terpstra statistics for all five metrics.

\begin{table}[htbp]
\centering
\caption{Geometry vs.\ rung order: full statistics for all five metrics (exact-permutation tests, Section~\ref{sec:stats}).}
\label{tab:s1-geometry-stats}
\small
\begin{tabular}{lcccccc}
\toprule
Metric & $\tau$ (full, $n=13$) & $p$ (exact) & $\tau$ (common-seed, $n=9$) & $p$ (exact) & $J$ & $p_{JT}$ (exact) \\
\midrule
Condition number             & $0.840$  & $2.8\times10^{-5}$ & $0.866$  & $0.0012$ & $55.0$ & $1.4\times10^{-5}$ \\
Fisher ratio (HL)            & $0.168$  & $0.521$            & $-0.096$ & $0.831$  & $33.0$ & $0.261$ \\
Fisher ratio (scalar)        & $0.076$  & $0.798$            & $-0.289$ & $0.388$  & $30.0$ & $0.399$ \\
Mardia's kurtosis ($b$)      & $-0.137$ & $0.608$            & $-0.160$ & $0.668$  & $23.0$ & $0.739$ \\
Mardia's kurtosis ($z$)      & $-0.168$ & $0.521$            & $-0.160$ & $0.668$  & $22.0$ & $0.780$ \\
\bottomrule
\end{tabular}
\end{table}

\subsection*{S2. Full geometry-vs-AUROC statistics}

\begin{figure}[htbp]
\centering
\includegraphics[width=\textwidth]{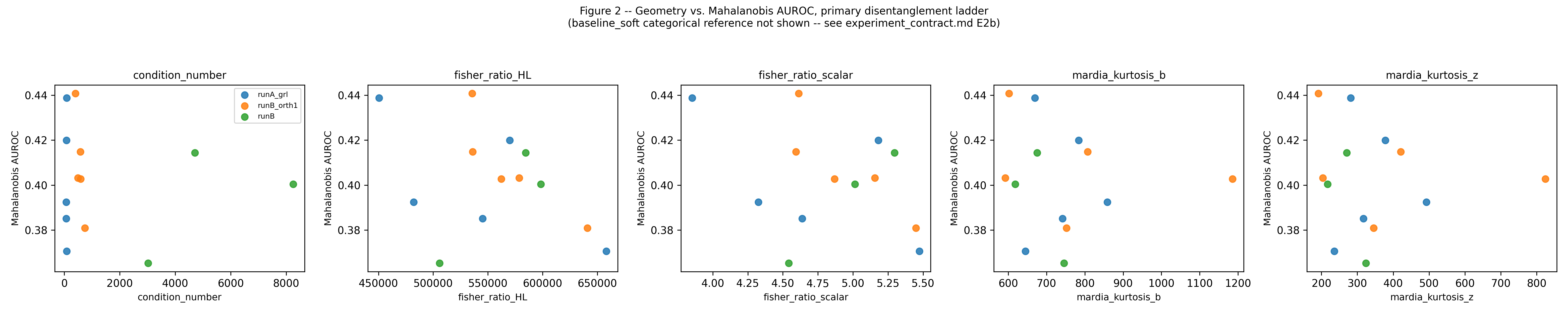}
\caption{All five geometry metrics vs.\ Mahalanobis AUROC.}
\label{fig:s2}
\end{figure}

Extends main-text Figure~\ref{fig:geometry-vs-auroc}, which shows only the condition-number panel. Figure~\ref{fig:s2} shows all five geometry metrics plotted against Mahalanobis AUROC, one point per checkpoint, colored by rung. Table~\ref{tab:s2-auroc-stats} reports the corresponding statistics, including a $10^5$-draw Monte Carlo cross-check of the exact Kendall's $\tau$ $p$-values.

\begin{table}[htbp]
\centering
\caption{Geometry vs.\ Mahalanobis AUROC: full statistics for all five metrics ($n=13$, exact-permutation Kendall's $\tau$; Monte Carlo cross-check at $10^5$ draws in parentheses).}
\label{tab:s2-auroc-stats}
\small
\begin{tabular}{lcc}
\toprule
Metric & $\tau$ & $p$ (exact / MC) \\
\midrule
Condition number       & $-0.128$ & $0.590$ / $0.589$ \\
Fisher ratio (HL)      & $-0.282$ & $0.204$ / $0.202$ \\
Fisher ratio (scalar)  & $-0.154$ & $0.510$ / $0.509$ \\
Mardia's kurtosis ($b$) & $-0.128$ & $0.590$ / $0.590$ \\
Mardia's kurtosis ($z$) & $-0.128$ & $0.590$ / $0.592$ \\
\bottomrule
\end{tabular}

\vspace{0.3cm}
\small AUROC itself vs.\ rung order (Jonckheere--Terpstra): $J=25.0$ (null mean $27.5$), $p_{\mathrm{exact}}=0.650$.
\end{table}

\subsection*{S3. Baseline reference: implementation detail}

The \texttt{baseline\_soft} checkpoint (main text, Section~\ref{sec:results-baseline}) is reported as a descriptive reference, not a fourth ladder point. Table~\ref{tab:s3-baseline} gives its full distance-scorer and probe AUROC (single checkpoint, seed 42). Its geometry metrics (condition number, Fisher ratio, Mardia's kurtosis) were not computed: because \texttt{baseline\_soft} was never load-bearing for any statistical claim in this study, computing its geometry was out of scope.

\begin{table}[htbp]
\centering
\caption{\texttt{baseline\_soft}: distance-scorer and domain-probe AUROC ($n=1$ checkpoint).}
\label{tab:s3-baseline}
\begin{tabular}{lc}
\toprule
Scorer / probe & AUROC \\
\midrule
Mahalanobis          & 0.863 \\
Cosine-to-centroid   & 0.715 \\
$k$-NN ($k=1$)       & 0.865 \\
$k$-NN ($k=10$)      & 0.853 \\
$k$-NN ($k=50$)      & 0.844 \\
Distance-scorer mean & 0.828 \\
\midrule
Probe: logistic regression & 0.998 \\
Probe: linear SVM          & 0.997 \\
Probe: random forest       & 0.977 \\
\bottomrule
\end{tabular}
\end{table}

\subsection*{S4. ID/OOD Mahalanobis distance distributions}

\begin{figure}[htbp]
\centering
\includegraphics[width=\textwidth]{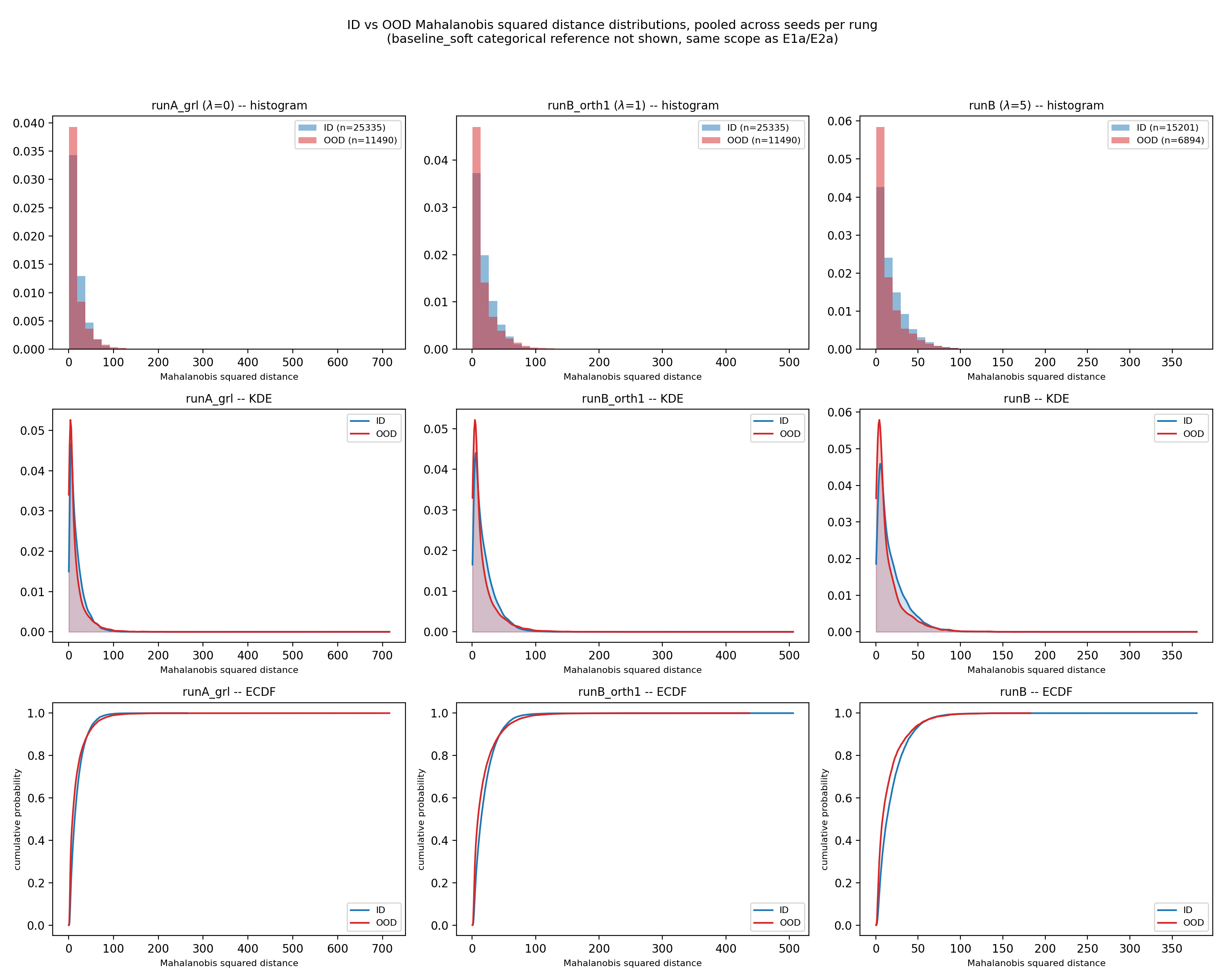}
\caption{ID vs.\ OOD Mahalanobis squared-distance distributions, all three rungs.}
\label{fig:s4}
\end{figure}

Figure~\ref{fig:s4} shows the full histogram, kernel density estimate, and empirical CDF of the ID (ISIC-test) and OOD (PAD-UFES) Mahalanobis squared-distance distributions, pooled across seeds within each rung, supporting the median-gap numbers reported in main-text Section~\ref{sec:results-distance-gap}. The distributions are visibly overlapping, not cleanly separated.

\subsection*{S5. Full nearest-centroid confusion matrix}

\begin{figure}[htbp]
\centering
\includegraphics[width=0.85\textwidth]{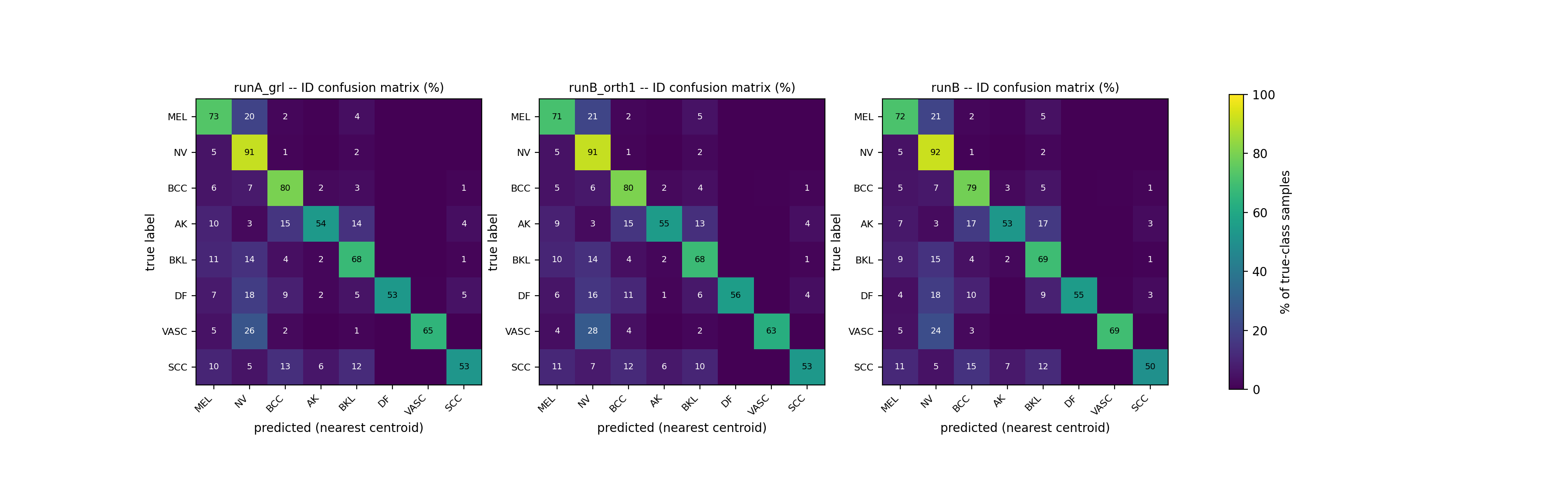}
\caption{Full ISIC-test nearest-centroid confusion matrix, all 8 classes.}
\label{fig:s5}
\end{figure}

Figure~\ref{fig:s5} shows the complete $8\times8$ nearest-centroid confusion matrix for ISIC-test samples (true label vs.\ predicted centroid), supporting the majority-class-only summary in main-text Figure~\ref{fig:mechanism}B.

\subsection*{S6. Full predicted-class distribution for PAD-UFES}

\begin{figure}[htbp]
\centering
\includegraphics[width=0.85\textwidth]{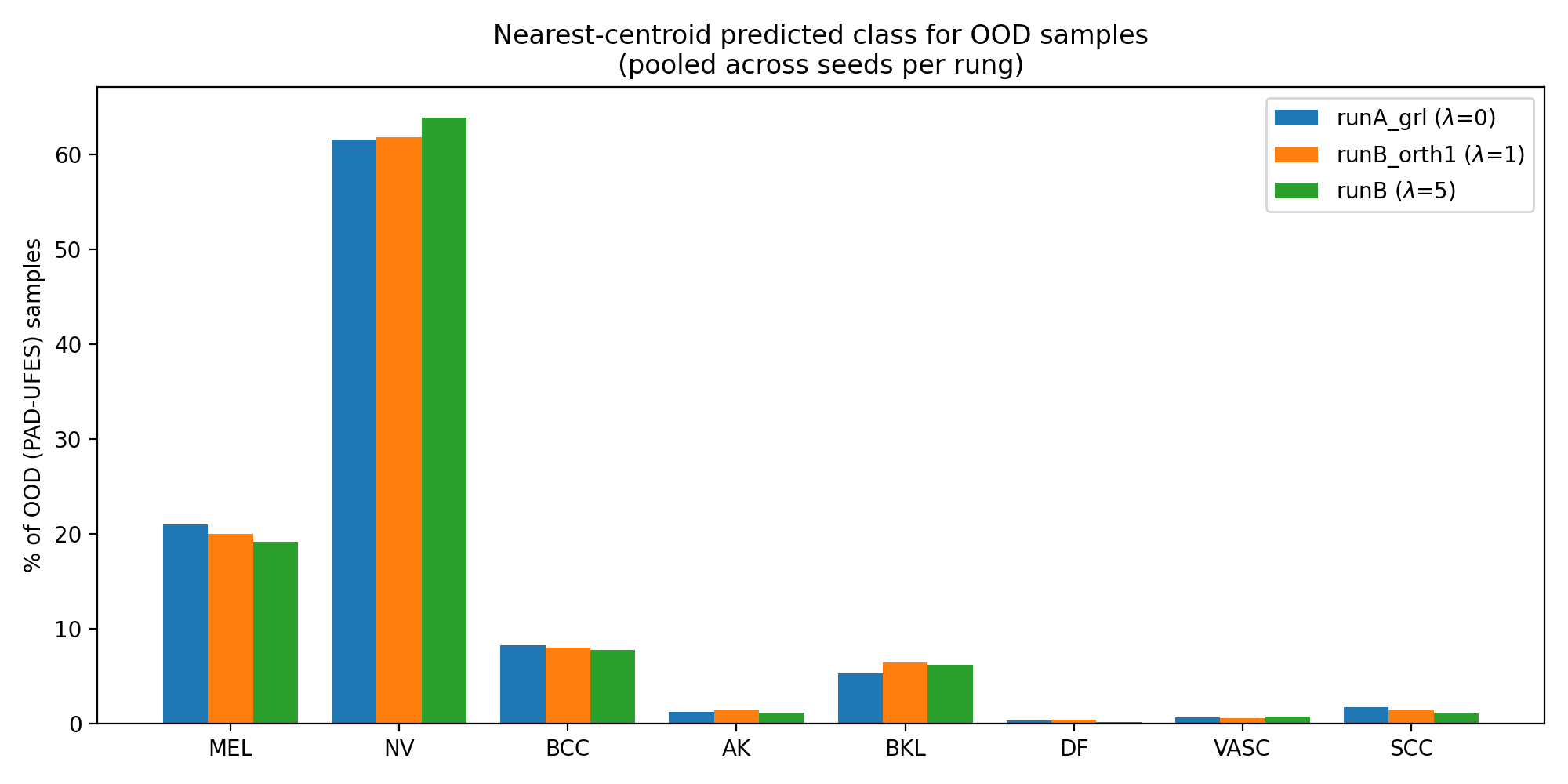}
\caption{Full PAD-UFES nearest-centroid predicted-class distribution, all 8 classes.}
\label{fig:s6}
\end{figure}

Figure~\ref{fig:s6} shows the complete nearest-centroid predicted-class distribution for PAD-UFES samples across all 8 ISIC classes, extending main-text Figure~\ref{fig:mechanism}B's majority-class (Nevus)-only summary.

\subsection*{S7. Full statistical test results: scorers and probes vs.\ rung order}

Table~\ref{tab:s7-scorer-probe-stats} reports exact-permutation Kendall's $\tau$ (and, for the five distance-based scorers, Jonckheere--Terpstra) for AUROC vs.\ rung order, for every scorer variant (main-text Section~\ref{sec:results-scorer-comparison}) and every probe (main-text Section~\ref{sec:results-probe}) individually. No metric or scorer reaches significance at $n=13$; none was omitted from this table.

\begin{table}[htbp]
\centering
\caption{AUROC vs.\ rung order, per scorer and per probe ($n=13$, exact-permutation tests).}
\label{tab:s7-scorer-probe-stats}
\small
\begin{tabular}{llccc}
\toprule
& Scorer / probe & $\tau$ & $p$ (Kendall, exact) & $p_{JT}$ (exact) \\
\midrule
\multirow{5}{*}{Distance-based}
& Mahalanobis        & $-0.076$ & $0.798$ & $0.650$ \\
& Cosine-to-centroid & $0.321$  & $0.194$ & $0.097$ \\
& $k$-NN ($k=1$)     & $-0.046$ & $0.898$ & $0.601$ \\
& $k$-NN ($k=10$)    & $0.015$  & $1.000$ & $0.500$ \\
& $k$-NN ($k=50$)    & $0.076$  & $0.798$ & $0.399$ \\
\midrule
\multirow{3}{*}{Non-distance-based}
& Energy             & $0.443$  & $0.065$ & $0.032$ \\
& ViM                & $0.076$  & $0.798$ & $0.399$ \\
& Density (KDE)      & $0.015$  & $1.000$ & $0.500$ \\
\midrule
\multirow{3}{*}{Domain probes}
& Logistic regression & $0.260$  & $0.302$ & --- \\
& Linear SVM          & $0.260$  & $0.302$ & --- \\
& Random forest       & $-0.107$ & $0.701$ & --- \\
\bottomrule
\end{tabular}
\end{table}